# Andreev spectroscopy of iron-based superconductors: temperature dependence of the order parameters and scaling of the $\Delta_{L,S}$ with $T_C$

T. E. Kuzmicheva, S. A. Kuzmichev, M. G. Mikheev, Ya. G. Ponomarev,
S. N. Tchesnokov, V. M. Pudalov, E. P. Khlybov, N. D. Zhigadlo

1. Introduction

Unexpected discovery in 2006 [1] of the first layered Fe-based high-temperature superconductor (HTSC) — LnOFePn (where Ln — lanthanide, Pn — pnictide; hereafter referred to as "1111") becomes a key issue of modern solid state physics. Since 2008, the class of iron-based superconductors has much expanded: several families of iron pnictides and chalcogenides have been synthesized [2–4]. The crystal structure of oxypnictides recalls that of cuprates and is in fact a stack of the superconducting Fe–As layers alternating along the $c$-direction with spacers, the nonsuperconducting oxide blocks, Ln–O. In spite of the pronounced layered structure and anisotropic physical properties, the electron subsystem in Fe-based superconductors is less quasi-two-dimensional in comparison with that in cuprate HTSC, because the height of the Fe–As blocks exceeds the thickness of the Cu–O planes, whereas the distance between superconducting blocks in iron-based superconductors is significantly less than that in cuprates. The latter seems to be a reason [5] why the critical temperature of Fe-based superconductors, though being as high as $T_C \approx 57.5$ K [6] still does not reach the cuprate one.

Superconductivity in novel materials emerges with the suppression of spin density wave ground state under doping of the superconducting Fe–As layers or under external pressure [7]. The key distinction from cuprates, however, is the multiband nature of newly-discovered superconductivity in iron-based materials. Band structure calculations showed (for a review, see [8]) the coexistence of the electron and the hole quasi-two-dimensional bands in these compounds, whereas the Fermi surfaces consist of slightly warped along the $c$-direction cylinders (near Γ and M points), where several superconducting condensates could form.

Contrary to the observation of the strong isotope effect [9], an early theoretical study showed [10] that the high temperature superconductivity in Fe-based compounds could not to be based on the electron-phonon interaction solely. Although the latter plays an important role, it is incapable [10] to explain the observable values of $T_C$ in the framework of the Eliashberg theory [11]. Taking into account the nesting of the Fermi surface sheets along the Γ–M-direction, the vicinity of the antiferromagnetic state [12,13], and the appearance of the experimentally observed [14] peak of dynamic spin susceptibility ("magnetic resonance"), Mazin et al. suggested the theoretical description of the mechanism of superconductivity in iron pnictides and chalcogenides based on sign-changing (in different bands) order parameter — the so-called $s^{\pm}$-model. The simplest initial model considers two isotropic order parameters (in the electron and the hole bands, correspondingly), which have equal amplitudes but are in antiphase (i.e. having opposite signs). The strong interband interaction mediated by spin fluctuations along the Γ–M-direction plays the key role in this model, whereas the intraband electron-phonon coupling is an order of magnitude weaker. Later, the initial model was found to be inadequate, in particular, because of its instability with respect to the impurity scattering [15–21], and the $s^{\pm}$-model has been extended [22].

According to the results of calculations in the framework of $s^{\pm}$-system, the magnetic resonance energy should not exceed the width of the large superconducting gap: $E_{res} \leq 2\Delta_L$ [13,14]. However, the experimentally observed (see [16] and the references therein), the susceptibility peak is rather smeared, whereas its position does not always satisfy the resonance condition [15,16]. Moreover, the most recent study for cuprates showed [23] that similar "magnetic resonance" is defined by the superconductor itself, while its energy coincides with $2\Delta$. Alternative theory of the two-gap state of Fe-containing superconductors is based on the coupling by orbital fluctuations [16]. This model is capable to describe such superconductivity within the framework of the constant-sign order parameter ($s^{++}$-type of symmetry).

Nevertheless, the two competing theories [12,16] do not deny the importance of the electron-phonon coupling underlying the intraband interaction within each of the superconducting condensates. On the other hand, as was shown in Ref. [24], the Allen—Dynes formula fails to provide such high $T_C$ values as experimentally obtained. This led the authors of Ref. [24] to conclusion on the significance of non-phonon mechanisms of the Cooper pairing for such compounds (including LnOFePn). Besides, the $T_C$ variation for various Fe-based superconductors was shown to be related to the variation of the density of states at the Fermi level. Within a single family of Fe-based superconductors, such $T_C$ variation can be caused by the modification of the chemical composition of spacers. Similar remote doping (δ-doping) of superconducting blocks is successfully practiced for cuprates. This work is dedicated to both, the experimental test of the aforementioned issue, and the comparison of the relative parameter, the characteristic ratio of the Bardeen—Cooper—Schrieffer (BCS) theory, $2\Delta_L/k_BT_C$, that evidences the strength of electron-boson coupling in LnOFeAs compounds (1111 family) with various $T_C$.

2. Review of the experimental studies

Despite the iron-based superconductivity has been discovered eight years ago, many of its aspects are still controversial. There is no common agreement on such issues as the mechanism of superconductivity, the number and the symmetry of superconducting gaps. The experimentally determined superconducting energy gaps and the magnetic resonance energy, the key parameters of iron-based compounds, are rather contradictory. Experimentalists face with a number of obstacles. The majority of traditional techniques used to determine the superconducting gap values (specific heat, London penetration depth, and nuclear magnetic resonance (NMR)) include processing of the experimental data on the basis of theoretical models. These models usually neglect an anisotropy of the transport and superconducting properties, thus giving the volume-averaged energy parameters. In particular, ~~the~~ synthesis of the large size oxypnictide single crystals is not still developed [see [25] and Refs. therein], therefore, their properties are measured mostly with polycrystalline samples. All this is the plausible reason of a low-resolution of the nonlocal techniques, whereas the measured parameters often appear to be reduced.

Indeed, temperature dependence of the London penetration depth measurements [26–28] reveal only a single superconducting gap with $2\Delta/k_BT_C \leq 3.52$, where the maximal ratio $2\Delta/k_BT_C \approx 3.4$ was obtained for single crystals [27]. Similar BCS-ratios were found in optical studies [29], specific heat [30] and NMR [31]. Two distinct superconducting gaps with rather high BCS-ratios, $2\Delta_L/k_BT_C \approx 7 \div 8$, have been resolved by NMR [32,33], and the authors do not exclude a possible d-wave symmetry.

The widely used local techniques, such as point-contact spectroscopy, scanning tunneling microscopy (STM), and the angle-resolved photoemission spectroscopy (ARPES) are based on probing surface properties of the sample, hence the data become strongly affected by the surface defects. In case of the ~~so-called~~ 1111 oxypnictides, which are easily cleaved along the ab-planes (between Fe–As and Ln–O layers) due to substantial anisotropy, their cleaved surface appears to be charged [34]. Alike effect is absent only for the (Li,Na)FeAs family (so-called 111 system), which could make this system a prime subject for surface techniques. However, there are additional principle obstacles: the alkali-metal atoms make the 111 superconductors extremely reactive, and, therefore, fast-degrading (particularly, in the presence of oxygen or water vapors).

To date, there are no ARPES data resolving both superconducting gaps. The ARPES studies certainly resolve clearly the large gap opening at the hole cylinder near the Γ-point of the Fermi surface, although, its anisotropy is ambiguous. As was demonstrated in Ref. [35] for Nd-1111 single crystals, the large gap has no nodes in the *k*-space, whereas its BCS-ratio $2\Delta/k_BT_C \approx 6.6$ being typical for a strong electron-boson coupling is close to the value estimated in our studies. At the same time, the data for La-1111 polycrystals [36] are in agreement with both, d- and s-wave models, while the BCS-ratio is lower: $2\Delta/k_BT_C \sim 3.6$ in the s-wave model approximation, and $2\Delta/k_BT_C \sim 4.1$ in case of the d-symmetry of the order parameter.

Dynamic conductance of the tunnel NIS-contacts (N — normal metal, S — superconductor, I — insulator) was studied in a number of works by STM [37–41]. These studies have resolved the only superconducting gap with $2\Delta/k_BT_C = 3.5 \div 4$. In Refs. [42,43] NIS-spectra containing peculiarities cased by two isotropic gaps $\Delta_L$ and $\Delta_S$ were obtained with Nd-1111 polycrystalline samples. The BCS-ratio for the large gap was estimated as $2\Delta_L/k_BT_C = 6.2 \pm 0.7$, similar to that obtained by NS-spectroscopy [44], as well as the results presented here. The presence of reproducible fine structure in the spectra was also shown in Ref. [43]. This structure corresponds well to the calculated Eliashberg function $\alpha^2F(\omega)$ and the phonon density of states obtained experimentally in Ref. [45]; therefore it is believed to have phonon nature [43]. Moreover, the critical temperature

$T_C^{theor}$ = 48.8 K estimated from $\alpha^2 F(\omega)$ calculations using the single-gap s-wave model was found similar to the real critical temperature of the sample $T_C^{sample}$ ≈ 51 K, thus confirming the conclusion of Ref. [43] on the intraband strong electron-phonon coupling.

Impressing is the divergence of the experimental results on the superconducting gap values (more than a factor of spread of the $2\Delta_L/k_BT_C$ values: $2\Delta_L/k_BT_C = 3.5 \div 22.3$), and on the number of the gaps observed [see reviews 46,47]. This divergence is intrinsic to the most popular tunneling technique such as point contact spectroscopy of NS-junctions. Unfortunately, in case of the 1111 polycrystals, the technique faces with such problems as degraded surface or inability of silver paste use for the point contact. The dynamic conductance spectra obtained in Refs. [48,49] with La-1111 and Sm-1111 contain pronounced maxima related with the small gap, whereas the large gap peculiarities are faded. Strong asymmetry of the NS-spectra and the necessity in seven parameter fitting (for two-gap superconductor) surely obstruct the data interpretation and lead leading to the data spread. For example, the BCS-ratio [48] for Sm-1111 with various dopant concentrations and the corresponding range $13.5 \leq T_C \leq 52$ K varies within the range $8 \div 22$ for the large gap, and $1.9 \div 6.8$ for the small gap. Such great variations could be caused only by changing of the coupling mechanism within the doping range.

The latter assumption, especially for 1111 family, seems highly improbable. The gap temperature dependencies measured in Ref. [49] for La-1111 and Sm-1111 are strongly dissimilar. For Sm-1111 samples, the BCS-like "closing" of both gaps was observed [49]. At the same time, for La-1111, the large gap vanished abruptly already at ~2/3$T_C$, whereas the small gap deviated from the BCS-like curve and linearly tended to zero at $T_C$ [49]. Such behavior of the gaps does not agree with the Moskalenko and Suhl system of gap equations [50–52] for two-gap superconductivity and was not confirmed in any subsequent works with Fe-based superconductors. The authors of Ref. [49] suggest no explanation to that gap behavior also, referring to as an "artefact". Summarizing some experimental data in Ref. [49], this group of researchers presented the dependencies of the BCS-ratios (for the large and the small gap) on critical temperature for various iron-based superconductors. According to Ref. [49], $2\Delta_{L,S}/k_BT_C(T_C)$ increases with $T_C$ decreasing, dramatically rising at $T_C \to 10$ K. The latter seems us unconvincing, both from the common sense arguments and because is unjustified by the experiment.

3. Experimental details

The review presented in Sec. 2 demonstrated the necessity of *direct* highly precise experimental probing of the superconducting gaps and their temperature dependencies to clarify features and the mechanism of the iron-based superconductivity. To address this issue, we used Andreev spectroscopy [53,54] of the *symmetric* SnS-contacts. In such contact when its diameter *a* is less than the quasiparticle mean free path *l* [55] an effect of multiple Andreev reflections from both NS-boundaries occurs. In this case, the current-voltage characteristic (CVC) shows a linear region with an excess current at low bias ("foot") and series of peculiarities at bias voltages

$V_n = 2\Delta/en$  ($n = 1, 2,\ldots$),     (1)

referring to as the subharmonic gap structure (SGS) [56–59]. In case of high-transparent contact, the dynamic conductance spectrum dI(V)/dV shows the series of *minima*. For a two-gap superconductor, two such SGS corresponding to both gaps should appear [60,61]. A finite *l/a* ratio at low biases *V* decreases the probability of ballistic transport through the SnS-interface for the quasiparticle, which causes fading of the Andreev reflection peculiarities with the harmonic number *n*. When *V* decreases (and *n* increases) to satisfy $an \approx l$, the excess current saturates. As bias tends further to zero, the current decreases. The latter elucidates the linearity of the "foot" in CVC.

In our studies we used the break-junction technique [62] based on the formation in the bulk of the sample of twin touching cryogenic clefts separated by the weak link (the so-called ScS-contact, where *c* is the constriction). The sample made as a thin rectangular plate of about $3 \times 1.5 \times 0.2$ mm$^3$ was mounted on a springy holder using a standard four-terminal connection. To fix the sample, we used liquid pads of In-Ga alloy at its corners. Then the sample holder was cooled down to T = 4.2 K, and deformed gently enough to generate a microcrack separating the sample into two superconducting banks. The I(V) and dI(V)/dV characteristics of our break-junctions in Fe-based superconductors [63–75] are typical for high-transparency Andreev regime with a ballistic transport (see Figs. 2b and 5 from [57], Fig. 4 from [59]).

The number of grains cleaved during the formation of the cryogenic cleft is known to be dependent on the $P_{grains}/P_{layers}$ ratio ($P_{grains}$ is the intergrain, and $P_{layers}$ — the interlayer solidity). Therefore, one can expect a high ratio of the cleaved and not cleaved crystallites along the microcrack for an annealed polycrystalline samples. Simple calculations show $P_{grains}/P_{layers} = 1.1$ is sufficient to expect about 6 % of the cleaved grains, whereas $P_{grains}/P_{layers} = 2.5$ would cause the cleavage of every second crystallite. The latter provides successful application of the break-junction technique for both, single and polycrystalline samples of layered compounds [71].

Due to the pronounced layered structure of the iron-based superconductors, the microcrack passes along the ab-planes and forms steps-and-terraces on the cryogenic clefts. The height of such steps is obviously a multiple of the *c* lattice parameter. The steps-and-terraces can be realized not only as single ScS-contacts, but as array structures of the S-c-S-c-…-S type. For Fe-based superconductors, the Fe–As blocks act as "S", whereas Ln–O spacers act as weak links. Obviously, the array contact is electrically equivalent to a sequence of identical ScS-contacts. Therefore, the bias voltage for any peculiarities (caused by bulk properties of the sample) in CVC and the dynamic conductance should be *N* times larger (where *N* is the number of junctions in the stack) for array contact in comparison with that of single ScS-contact. Such array contacts were for the first time observed on cuprates [76–79], and then on other layered superconductors [67–72,74,75]. In the Andreev mode (i.e. when the weak link formally acts as the normal metal), such array demonstrate the intrinsic multiple Andreev reflections effect (IMARE) [78], similar to the intrinsic Josephson effect [76,78,80]. Therefore, the gap value may be determined using the formula $V_n = 2\Delta \cdot N/en$.

Were one trying to attribute the observed S-c-S-c-…-S contacts in break-junction experiments to formation of a sequence of crystallites touched via intergrain boundaries (rather than to the bulk effects on natural stack structures), the result would not stand up to criticism. In the former case, due to inequality of the intergrain boundaries, the position of the main gap peculiarities would be *accidental* rather than multiple to $2\Delta/e$. Also, the shape and the fine structure of peculiarities in dynamic conductance spectra would be *irreproducible* under a mechanical readjustment of the contact. Moreover, increasing the number $N$ of consequently connected grains (having different resistance in the normal state) along with the number of inequivalent intergrain boundaries would cause dramatic smearing of peculiarities in the dI/dV-spectrum. In our studies, the opposite is observed: when $N$ increases, the spectra quality improves, whereas the position and the shape of SGS peculiarities are *well-reproduced* under scaling the bias axis by natural $N$. Furthermore, we observe similar tendency for single crystals of various superconductors.

Summarizing the above, we conclude on a number of advantages of the employed break-junction technique:

(a) the presence of clean cryogenic clefts in the bulk of the sample: under the microcrack formation, the superconducting banks can precisely slide on top of each other, along the ab-planes. The microcrack is not opened preventing degradation of the surface;

(b) the break-junction provides local probing of the superconducting parameters (within the contact area of about several nm in diameter);

(c) the applicability for both, single crystals, and polycrystalline samples of layered compounds. In particular, in Ref. [71] we showed the average diameter of the break-junction contact, 10–30 nm, to be several tens smaller than both, the crystallite size, and typical terrace width in the samples under study;

(d) the opportunity of probing several tens of single and array contacts on cryogenic clefts of one and the same sample, using a mechanical readjustment; this improves data statistics, and enables to conclude on reproducibility of experimental data and nanoscale homogeneity of the samples. The latter ability of the break-junction technique is similar to that of STM;

(e) the break-junction technique prevents the contact point from an overheating due to remote current leads (a true four-point connection);

(f) the realization of both experimental techniques with iron-based superconductors: Andreev spectroscopy and intrinsic Andreev spectroscopy. The latter being implemented to high-quality natural array contacts (nearly unaffected by an overheating) guarantees measurements of the bulk superconducting gap values. A contribution of any surface defects to the dynamic conductance is reduced by $N$ times for such arrays; it also means the $N$ times increased precision of the superconducting parameters determination [71].

Summarizing, we note that the intrinsic Andreev spectroscopy realized by the break-junction technique is the only known tool for *local* probing of the *bulk* (irreduced) superconducting gap values. The gap magnitudes with this technique can be determined from dynamic conductance of the SnS-contacts *directly,* up to $T_C$ [56,57,59] with no need to fit the experimental spectra with several parameters.

4. Sample characterization

We used polycrystalline samples of oxypnictides based on different lanthanides, and both, single and polycrystalline samples of FeSe. Typical temperature dependencies of resistance near the superconducting transition are shown in Fig. 1. Nearly optimal $GdO_{1-x}F_xFeAs$ samples with nominal fluorine concentrations $0.09 \leq x \leq 0.21$ and bulk critical temperatures $T_C^{bulk} = 46 \div 53$ K, and $GdO_{0.88}FeAs$ (circles in Fig. 1) with $T_C^{bulk} \approx 50$ K have been made by high pressure synthesis detailed in Refs.[66,81]. $CeO_{0.88}F_{0.12}FeAs$ samples (up triangles in Fig. 1) with $T_C^{bulk} \approx 41$ K were synthesized similarly. $Sm_{1-x}Th_xOFeAs$ samples (down triangles and squares in Fig. 1) with the range of Th doping $0.15 \leq x \leq 0.3$ had the critical temperatures $T_C^{bulk} = 40 \div 52.5$ K, correspondingly; their synthesis and characterization are detailed in [82]. $LaO_{0.9}F_{0.1}FeAs$ samples (rhombs in Fig. 1) had the lowest critical temperature among the oxypnictides studied, $T_C^{bulk} = 21 \div 29$ K [63,83]. The bulk critical temperature of FeSe single crystals (solid line) was $T_C^{bulk} \approx 10$ K [73], and of polycrystalline samples (pentagons in Fig. 1) was $T_C^{bulk} \approx 11$ K [64,70]. All the $T_C^{bulk}$ were determined as the maximum of $dR(T)/dT$.

5. Experimental results

Figure 2 shows the dynamic conductance spectrum of single SnS-contacts (upper curve) and the normalized spectra of Andreev arrays (with $N = 2, 6, 9$, and 8). The contacts were formed in several optimally doped $GdO_{1-x}F_xFeAs$ and $GdO_{0.88}FeAs$ samples [65,66,71] with critical temperatures $T_C = 46 \div 50$ K. The spectra demonstrate pronounced peculiarities involving the most intensive minima at $V_{L1} \approx 22$ mV and $V_{L2} \approx 11$ mV (their position is marked in Fig. 2 by gray areas covering 10 % range of uncertainty, and by $n_L$ labels), as well as peculiarities at $V_{L3} \approx 7.3$ mV (labeled as $n_L = 3$). According to Eq.(1) for subharmonic gap structure (SGS), these peculiarities are interpreted as the first, second and third Andreev minima for the large gap $\Delta_L \approx 11$ meV. The next peculiarity located at lower bias $V \approx 5.5$ mV is much more intensive than the third minimum for $\Delta_L$, and, hence, cannot be attributed to the forth Andreev peculiarity of the large gap SGS. Therefore, this minimum starts the new SGS consisting of the peculiarities at $V_{S1} \approx 5.5$ mV and $V_{S2} \approx 2.7$ mV (marked by dashed areas depicting the uncertainty range, and by $n_S$ labels in Fig. 2), and determining the small gap value $\Delta_S = 2.7 \pm 0.3$ meV.

In our studies we have obtained several hundreds of various single and array Andreev contacts; in the corresponding spectra two distinct SGS have been observed [63–75]. The positions of peculiarities for both, the large and the small gaps coincided after the scaling of array contact spectra to those of the single contact (i.e. the scaling of the bias axis by an integer).

This results mean that we observe reproducibly multiple Andreev reflections effect (MARE) and the intrinsic multiple Andreev reflections effect (IMARE); the SnS-Andreev and intrinsic Andreev spectroscopy give similar results. Since the array contacts are a part of the natural structure of iron-based superconductors, we conclude on the Andreev-type transport along the *c*-direction in these compounds.

High quality of the break junctions enables us to resolve fine structure of the Andreev reflection peculiarities. It is worth of noting the reproducibility of slightly asymmetric shape of the first minima for the large gap (see Fig. 2): when bias decreases, the dynamic conductance firstly falls down abruptly, and then smoothly rises. The latter could be a sign of a 20–30 % anisotropy of $\Delta_L$ in the *k*-space (i.e., an extended s-type symmetry of the order parameter) [71]. Making analogous conclusion on the $\Delta_S$ anisotropy is impossible due to the location of the small gap SGS in the increased dynamic conductance region ("foot" of the large gap) which impedes an interpretation of $\Delta_S$ Andreev minima shape. The significant amplitude of the first SGS minima for both $\Delta_L$ and $\Delta_S$ gaps allows to conclude indirectly on the absence of nodes in the *k*- direction dependence of the superconducting order parameters.

Figure 3 shows temperature variation of the large and the small gap SGS in the normalized (by the integer *N*) dI(V)/dV-spectrum of an array contact. The corresponding spectrum measured at T = 4.2 K in $GdO_{0.88}FeAs$ sample is presented in Fig. 2 (the third curve from the top). The spectra were shifted along the vertical scale, for clarity. The strong smearing of $n_L = 3$ and $n_S = 2$ minima could be caused by relatively large contact diameter comparable to the mean free path. Nevertheless, the main gap peculiarities (marked as $n_L = 1$ and $n_S = 1$ in Fig. 3) are clearly seen up to $T_C$ thus allowing to obtain the temperature dependencies of the large and the small gap just using formula $2\Delta_{L,S}(T) = V_{nL,S}(T)\cdot en$ with no fitting procedure.

Noteworthy, the asymmetry of minima is invariable regardless of *T*. With the temperature increasing, these minima shift towards the low bias, and at $T \approx 50 \text{ K} = T_C^{local}$ the dynamic conductance becomes linear. This means the transition of the contact area (of about 10–30 nm in diameter) to the normal state. The local critical temperature of the contact can differ from the bulk one $T_C^{bulk}$ determined by non-local techniques, such as resistive measurements, etc. (see Fig. 1). Given $T_C^{local}$ of the contact is determined one can precise calculate precisely the BCS-ratio: for this contact, $2\Delta_L/k_B T_C^{local} \approx 5.2 \gg 3.52$, $2\Delta_S/k_B T_C^{local} \approx 1.3 \ll 3.52$. Similar measurements of the Andreev spectra within the range $4.2 \text{ K} \leq T \leq T_C^{local}$ give temperature dependence of both gaps in various iron-based superconductors. As an example, $\Delta_L(T)$ (solid symbols) and $\Delta_S(T)$ (open symbols) for Sm-based oxypnictide (triangles and squares) and Gd-based one (circles) [69–74] are presented in Fig. 4. Firstly, it is obvious that the large and the small gaps behave distinctly with temperature increasing. The large gap $\Delta_L(T)$ dependence, in general, resembles the standard BCS-type (dash-dotted lines in Fig. 4), but slightly sags down due to the interband interaction with the second condensate. The small gap $\Delta_S(T)$ deviates significantly from the BCS-like curve: at the temperatures of about *, the small gap lowers abruptly, and then tends smoothly to zero at the local critical temperature. This demonstrates the superconductivity to vanish simultaneously in both condensates at the common critical temperature $T_C^{local}$.

The different temperature behavior of the large and the small gaps, their discrepancy with the standard BCS-like function, and the reproducibility of the $\Delta(T)$ curves (not only for the various contacts in the same sample, but even for the contacts obtained on various compounds of the 1111 family) manifest the two-gap superconductivity (see Fig. 4), meaning that the peculiarities in SnS-spectra are related to the distinct SGS, and, therefore, describe properties of distinct condensates. Moreover, in accordance with our studies, the gap temperature dependencies look similarly for iron-based superconductors of various families (see Fig. 4, and Refs. [64,67–70,72,74,75]).

To analyze the measured temperature dependencies, we fitted them using two-gap system of equations by Moskalenko and Suhl [50,52] with renormalized BCS-integral (shown by the solid lines in Fig. 4). This system involves a set of four electron-boson interaction constants $\lambda_{ij} = V_{ij}N_j$, where $V_{ij}$ are the matrix elements of $k$-space interaction between $i^{th}$ and $j^{th}$ bands, $N_j$ is normal state density of states at the Fermi level of $j^{th}$ band. Figure 4 demonstrates a good correspondence of the theoretical curves to the experimental data, thus showing that the two-gap BCS-like model is sufficient to describe properties of the studied Fe-based superconductors.

Thereby, experimentally observed deviation of $\Delta_L(T)$ and $\Delta_S(T)$ from single-band dependencies could be caused by a $k$-space proximity effect [84] between two condensates due to the nonzero interband coupling. The fitting of the measured gap temperature dependencies enables us to compare the strength of interband and intraband interactions of the condensates for various members of the 1111 family.

6. Discussion

The dependencies of the large (solid symbols) and the small gap (open symbols) on the $T_C$ for the studied iron-based superconductors, such as Gd-1111, Sm-1111, Ce-1111, La-1111, and FeSe, are presented in Fig. 5. The critical temperatures of these compounds cover almost the whole range up to $T_C = 53$ K accessible for the 1111 family. The data related to the contacts with known $T_C^{local}$ are depicted by the large symbols. For contacts with the spectra measured only at $T = 4.2$ K we used the $T_C^{bulk}$ (corresponding data are shown by the small symbols). As expected, the data points related to $T_C^{bulk}$ appear to lie lower than those related to $T_C^{local}$, since, in general, $T_C^{bulk} > T_C^{local}$. The latter fact evidences for the necessity of using namely local superconducting parameters for correct estimation of $2\Delta_{L,S}/k_BT_C$. The use of bulk values makes doubtful any conclusions on whether the coupling mechanism is constant or mutable under variations of electron doping for both, 1111 family, and other superconductors.

Our data (see Fig. 5) shows that the two gaps are directly proportional to the critical temperature within the range $9\,K \leq T_C \leq 53\,K$. The observed scaling with $T_C$ means the ratio of the magnitudes of the large and the small gaps to be constant within these $T_C$. The averaging of the result of Fig. 5 leads to $\langle \Delta_L/\Delta_S \rangle \approx 4$. Note that the magnitude of the order parameters obtained on both single crystals and polycrystalline samples of FeSe are similar and belong to same group (down triangles in Fig. 5).

The BCS-ratios for each band are also unchanged within the uncertainty range (Fig. 6). For the large gap, the ratio is $2\Delta_L/k_B T_C = 4.6 \div 6.0$, notably exceeding the weak-coupling BCS-limit 3.52. This is probably caused by a strong electron-boson coupling in the bands with the large gap. The solid horizontal lines in Fig. 6 depict the average values of the BCS-ratios. Obviously, there is no tendency of significant variation of the interaction strength (and, therefore, to changing of the coupling mechanism), the average value of the BCS-ratio is $\langle 2\Delta_L/k_B T_C \rangle \approx 5.2$. Similar energy of magnetic resonance for oxypnictides $E_{res}/k_B T_C = 5.1 \div 5.3$ was obtained from polarized neutron scattering in [85,86]. For the small gap, $2\Delta_S/k_B T_C = 0.6 \div 2.0 \ll 3.52$. The obvious reason is that the observed common $T_C$ is not the "eigen" critical temperature of the small-gap condensate, and does not describe its properties.

The "eigen" parameters of each superconducting condensate (in a hypothetical case of zero interband interaction), such as, $2\Delta_L/k_B T_C^L$ and $2\Delta_S/k_B T_C^S$, together with the relative coupling constants (normalized to intraband $\lambda_{LL}$ for the condensate with the large gap), the partial densities of states ratio $\alpha = \lambda_{LS}/\lambda_{SL} \equiv N_S/N_L$ (in a case of zero Coulomb repulsion constants $\mu^*$), and the ratio of effective intraband coupling to interband one $\beta = \sqrt{\lambda_{LL} \cdot \lambda_{SS} / (\lambda_{LS} \cdot \lambda_{SL})} = \sqrt{V_{LL} \cdot V_{SS}}/V_{LS}$ have been estimated from fitting of the gap temperature dependencies with the extended model by Moskalenko and Suhl (see the Table). Noteworthy, all the $\Delta_{L,S}(T)$ dependencies measured for different families look similarly. For all studied compounds, the eigen $2\Delta_L/k_B T_C^L$ ratio for the large gap condensate is about $4.2 \div 4.8$, and for the small gap condensate on average is somewhat less, from the BCS-limit 3.5 to 4.5. The superconductivity in both condensates, therefore, can be described in the framework of the strong-coupling Eliashberg theory [11].

Taking into account the presence of the strong isotope effect in iron-based superconductors [9], and some experimental data of Refs. [19,43], we conclude on a strong electron-phonon interaction *within each band*. The intraband interaction in the small gap bands is weaker; $\lambda_{SS}$ is on average 60 % of $\lambda_{LL}$ which also supports the latter conclusion. In spite of intraband interaction plays the key role in superconductivity of oxypnictides ($\beta \gg 1$), for studying the two-gap state it is necessary to take also the interband interaction into account.

Our data show (see the Table) that the two condensates interact weakly, whereas namely due to $\lambda_{LS,LS} \neq 0$ the small gap does not vanish to zero until the local contact $T_C^{local}$; indeed, for zero interband coupling, according to the theory [50–52], the small gap would close at its eigen $T_C^S$. Therefore, for temperatures exceeding $T_C^S$, superconductivity in the $\Delta_S$-condensate is induced by the "driving" $\Delta_L$-condensate (i.e., due to the *k*-space proximity effect). The latter causes the large gap temperature dependence to bend down relative to the BCS-like function, leading to 20–30 % decreasing of the $T_C^{local}$ with respect to the eigen $T_C^L$ (see the Table). Since such sags are caused by the small gap band, their intensity should depend on the density of states ratio in two bands (the α parameter). We estimate the density of states in the small gap band to be of the order of magnitude higher than that in the large gap band.

We have shown the relative $\lambda_{ij}$ and the "eigen" BCS-ratios $2\Delta_i/k_B T_C^i$ to be nearly constant under variations of both, doping concentration, and constituting lanthanide Sm/Gd/Ce/La. The studied samples differ in the chemical compositions of the spacers rather than in superconducting Fe–As blocks. The degree of structural ordering of the superconducting Fe–As blocks is nearly unchanged, whereas the doping level varies leading to the density of states modification * in the bands.

Our studies thus prove the spacers in 1111 compounds to be charge reservoirs and are not involved directly in the superconductivity. The aforementioned variations of the chemical compositions are supposed not to affect the coupling mechanism and the strength of electron-boson coupling. In addition, taking into account similar quasi-two-dimensionality of both condensates and similarity of the Fermi surfaces for 1111 family and FeSe [87], the experimentally observed scaling of both gaps with the critical temperature becomes obvious, and agrees well with the theory [24].

7. Conclusions

We studied iron-based superconductors of various families with critical temperatures covering almost all range, from 9 K to maximal $T_C$ = 53 K. In natural arrays of contacts formed in these materials we observed intrinsic multiple Andreev reflections. The spectroscopy of such contacts demonstrated that the multiple Andreev reflections contribute significantly to the transport along the *c*-direction in iron-based superconductors. In all studied compounds we detected the two-gap superconductivity, determined the value of the large and the small superconducting gaps, and the corresponding BCS-ratios. The measured temperature dependencies of the large and the small gaps $\Delta_{L,S}(T)$ are similar for various families of the Fe-based superconductors and could be well-fitted in the framework of the two-band model by Moskalenko and Suhl. We concluded on the extended s-wave symmetry of the $\Delta_L$ order parameter (with an anisotropy of about 20–30 % in the *k*-space) and on the absence of nodes for $\Delta_S$.

Our studies showed that the BCS-ratio for the bands with the large gap $2\Delta_L/k_BT_C \approx 5.2$ is nearly constant within the whole range of $T_C$ (this means that coupling rate is unchanged), reflecting the ~20 % reduction of the $T_C^{local}$ in relation to the eigen $T_C^L$, and the large gap roughly corresponds to the energy of magnetic resonance $2\Delta_L \approx E_{res}$. This result requires a special theoretical consideration.

Our estimation of the relative coupling constants and eigen parameters of each condensate (in a hypothetical case of a zero interband interaction) $2\Delta_L/k_BT_C^L = 4.2 \div 4.8$ and $2\Delta_S/k_BT_C^S = 3.5 \div 4.5$ leads to indirect conclusion that namely a strong electron-phonon interaction in each condensate described in the framework of the Eliashberg theory [11] plays the key role in the superconductivity of iron-based oxypnictides. With it, the two condensates interact weakly with each other ($\beta \gg 1$, see the Table). Nevertheless, in comparison with another two-gap superconductor, $MgB_2$, for the studied Fe-based compounds the interband interaction appears to be stronger, whereas the intraband one — weaker. Indeed, for the σ-bands in $MgB_2$ the "eigen" BCS-ratio is about $2\Delta_L/k_BT_C^L \approx 5.1$ and $\beta = 10$–$20$ [88], while for Ln-1111 family $\beta = 5$–$15$.

According to our data, the mean value of the gap ratio is about $\Delta_L/\Delta_S \approx 4$ being nearly constant within the whole range $T_C = 9 \div 53$ K. The observed scaling of $\Delta_{L,S}$ with $T_C$, as was discussed above, is caused by changing of the density of states $N_{L,S}$ in the bands. According to the BCS theory, namely increasing of the intraband coupling constant $\lambda = VN$ enhances $\Delta$ and $T_C$. The ability of chemical doping to enhance the density of states $N$ is very limited for the known iron-based superconductors. As regards a perspective of increasing $V$, we point to a hypothetical possibility of boosting the intraband coupling by an increasing of phonon density of states at high energies. The latter could be realized, for example, by a variation of crystal lattice parameters of iron-based superconductors or by an implantation of light atoms with the unfilled $p$-electron shell (B, C, N) to the periodic structure of the spacer layers.

We thank Yu.F. Eltsev, A.V. Sadakov, K.S. Pervakov, L.F. Kulikova, P.I. Arseev, N.K. Fedorov, D.A. Chareev, A.N. Vasiliev, O.S. Volkova, T. Hänke, C. Hess, G. Behr, R. Klingeler, B. Büchner, Th. Wolf. The studies are supported by the Russian Foundation for Basic Research (projects nos. 13-02-01451, 14-02-90425).

| compound | Gd-1111 | | La-1111 | Sm-1111 | |
| --- | --- | --- | --- | --- | --- |
| $T_C^{local}$, K | 49 | 50 | 21 | 45 | 37 |

| $\Delta_L$, мэВ | 12.5 | 11.2 | 5.4 | 10.5 | 8.3 |
|---|---|---|---|---|---|
| $\Delta_S$, мэВ | 3 | 3 | 1.4 | 2.8 | 1.8 |
| $2\Delta_L/k_B T_C^{local}$ | 5.9 | 5.2 | 6 | 5.4 | 5.2 |
| $2\Delta_S/k_B T_C^{local}$ | 1.4 | 1.4 | 1.5 | 1.4 | 1.1 |
| $2\Delta_L/k_B T_C^L$ | 4.8 | 4.3 | 4.4 | 4.55 | 4.5 |
| $2\Delta_S/k_B T_C^S$ | 3.8 | 3.53 | 3.7 | 4.5 | 4.3 |
| $T_C^{local}/T_C^L$ | 0.81 | 0.83 | 0.73 | 0.84 | 0.87 |
| $\lambda_{SS}/\lambda_{LL}$ | 0.63 | 0.58 | 0.75 | 0.67 | 0.64 |
| $\lambda_{LS}/\lambda_{LL}$ | 0.26 | 0.37 | 0.36 | 0.18 | 0.18 |
| $\lambda_{SL}/\lambda_{LL}$ | 0.023 | 0.073 | 0.025 | 0.018 | 0.018 |
| $\alpha = \dfrac{\lambda_{LS}}{\lambda_{SL}} = \dfrac{N_S}{N_L}$ ($\mu^*=0$) | 11.2 | 5.1 | 15.5 | 10.3 | 9.7 |
| $\beta = \sqrt{V_{LL} \cdot V_{SS}}/V_{LS}$ | 10.4 | 4.6 | 9.4 | 14.4 | 14 |

Table 1. Parameters of the superconducting state estimated using $\Delta_{L,S}(T)$ fits for La, Sm, and Gd-based oxypnictides. Here $\lambda_{ij} = V_{ij} N_j$ are electron-boson coupling constants.

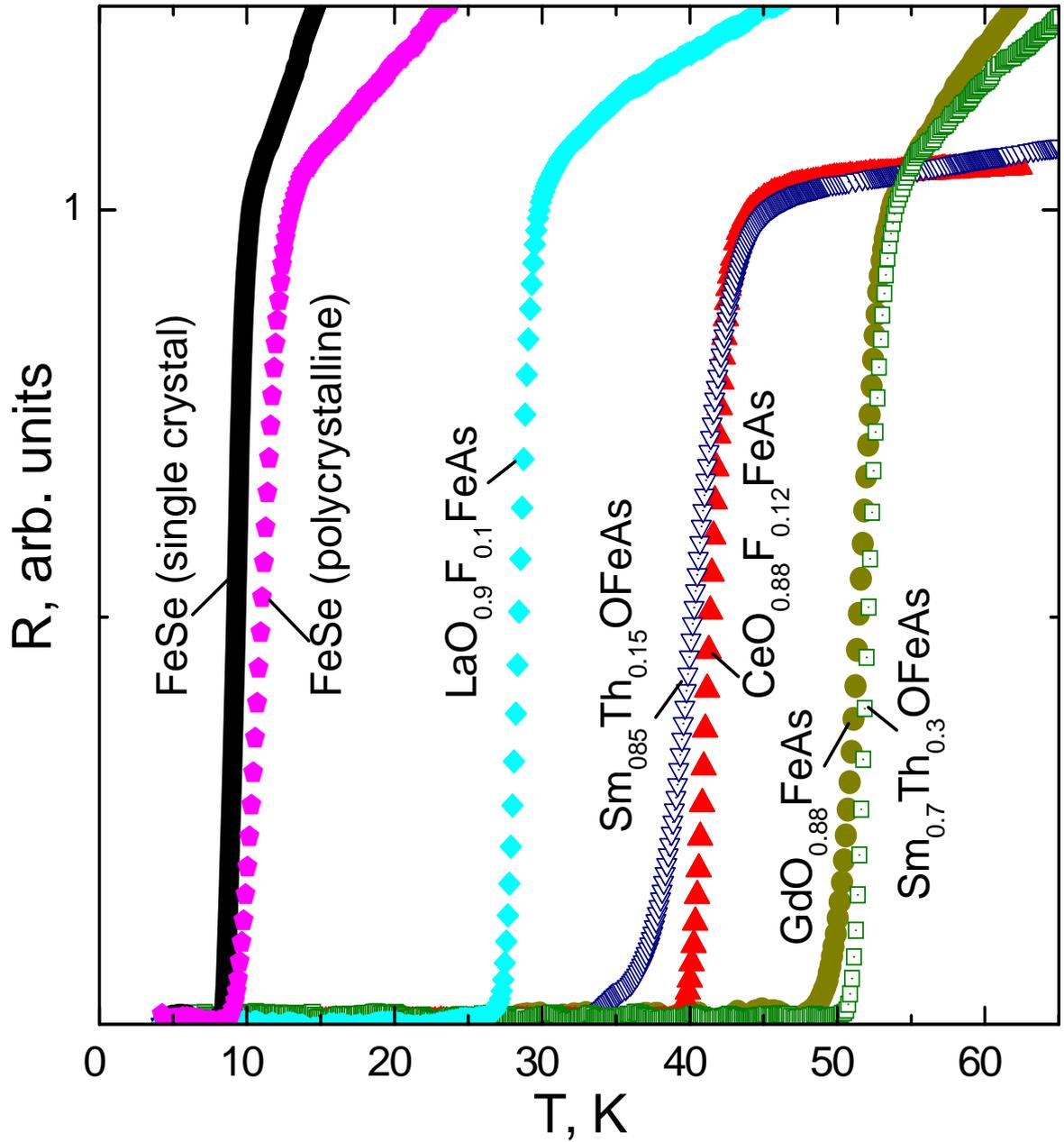

Fig. 1. Normalized temperature dependences of resistance in the vicinity of the superconducting transition for various Fe-based superconductors: FeSe single crystals (solid line) with bulk critical temperature $T_C^{bulk} \approx 9.5$ K, and polycrystals of FeSe (pentagons) with $T_C^{bulk} \approx 11$ K, $LaO_{0.9}F_{0.1}FeAs$ (rhombs) with $T_C^{bulk} \approx 28$ K, $Sm_{0.85}Th_{0.15}OFeAs$ and $Sm_{0.7}Th_{0.3}OFeAs$ (down triangles and squares) with $T_C^{bulk} \approx 40$ K and $T_C^{bulk} \approx 52.5$ K, correspondingly, $CeO_{0.88}F_{0.12}FeAs$ (up triangles) with $T_C^{bulk} \approx 41$ K, $GdO_{0.88}FeAs$ (circles) with $T_C^{bulk} \approx 51.5$ K.

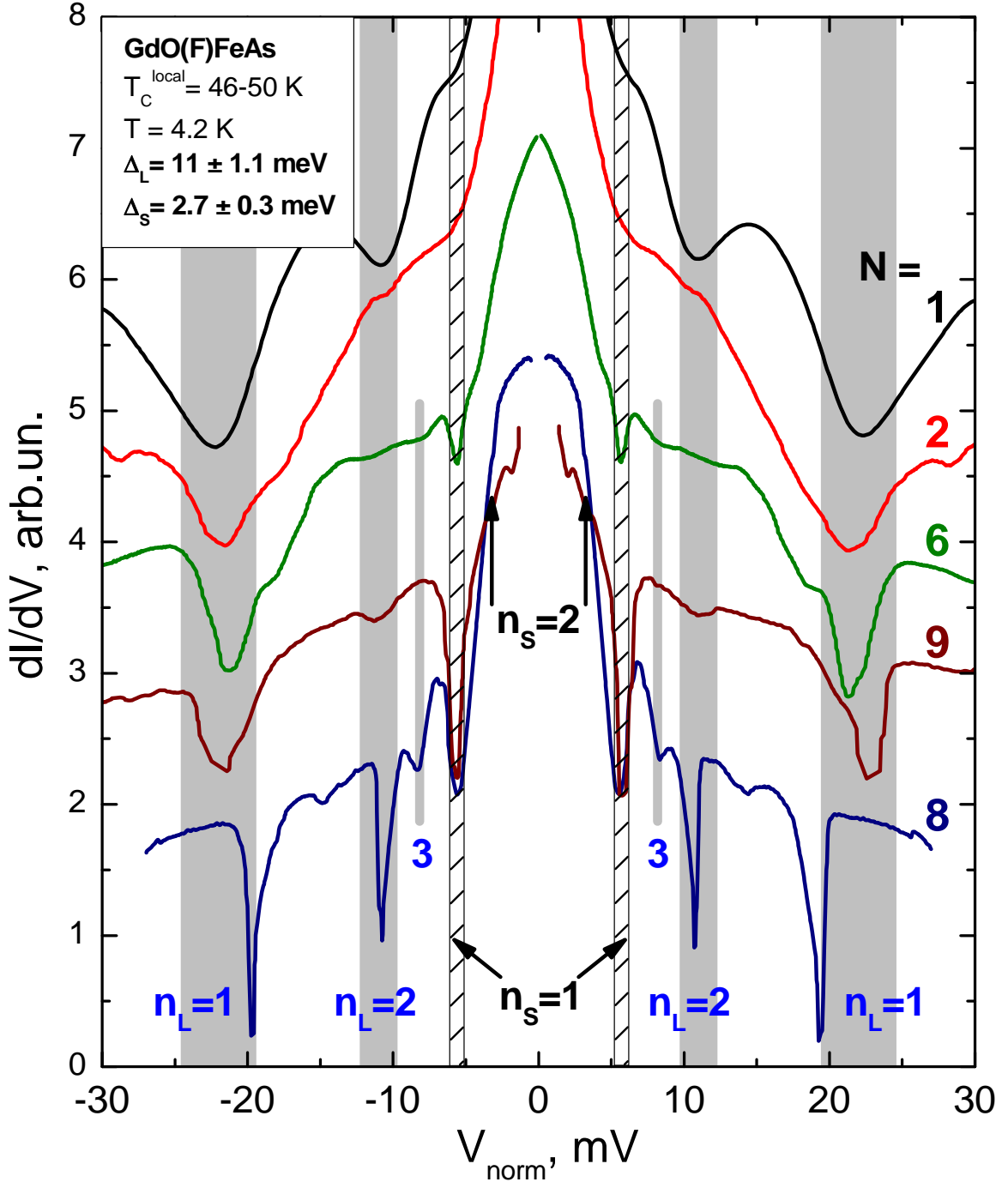

Fig. 2. (Color online) Dynamic conductance spectra for SnS-Andreev arrays (four bottom lines) with the number of junctions in the stack N = 2, 6, 9, and 8, correspondingly. The curves were normalized to the spectrum for single SnS-contact (top line, N = 1). The SnS-contacts were obtained on various GdO(F)FeAs samples with bulk critical temperatures $T_C = 46 \div 50$ K. The positions of subharmonic gap structure minima for the large gap $\Delta_L = 11.0 \pm 1.1$ meV are marked by gray areas and $n_L$ indexes, for the small gap $\Delta_S = 2.7 \pm 0.3$ meV — by dashed areas, arrows and corresponding $n_S$ indexes. The area width depict the range of gap values.

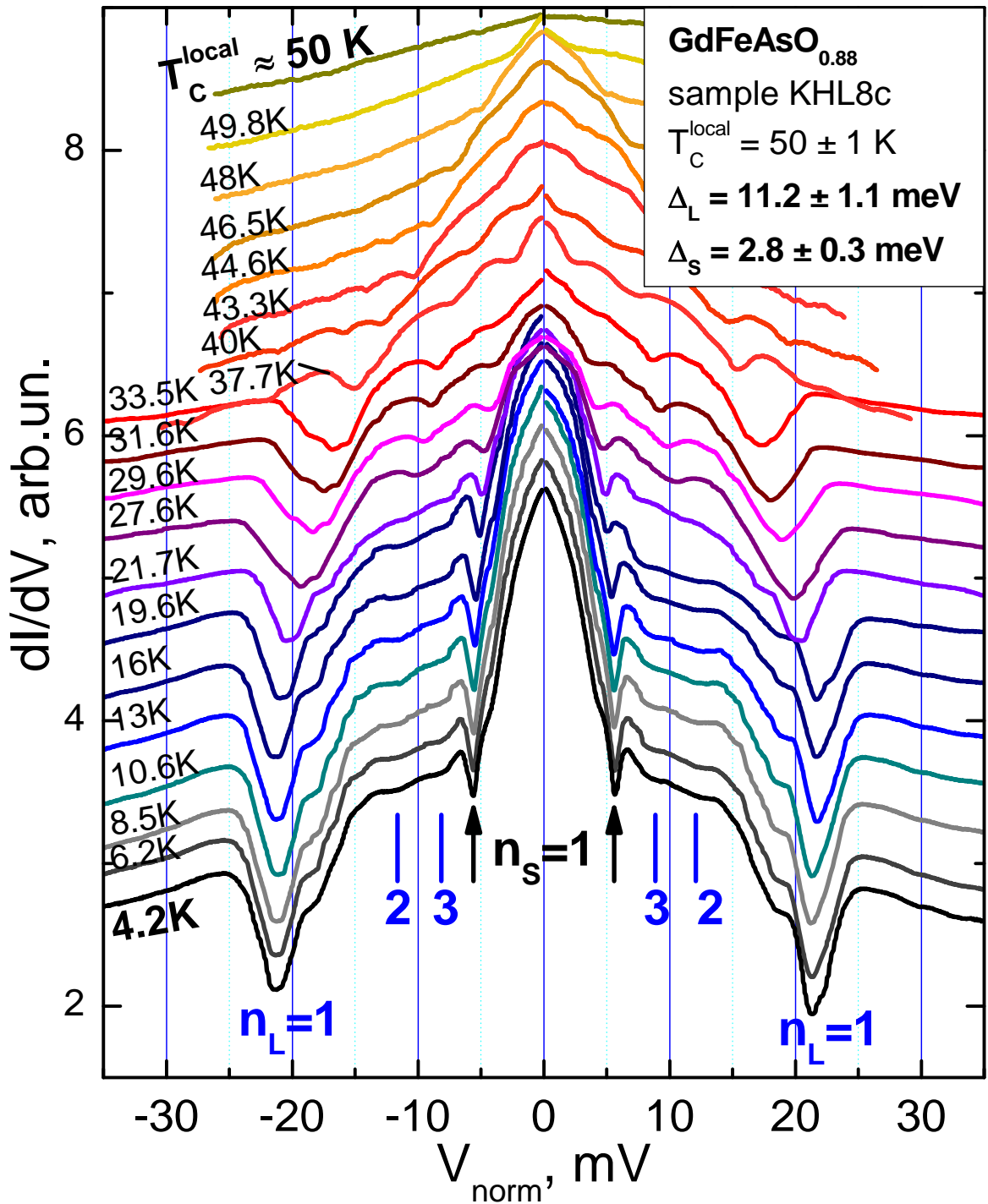

Fig. 3. (Color online) Dynamic conductance of SnS-contact in $GdO_{0.88}FeAs$ sample measured within the temperature range $4.2\ K \leq T \leq T_C^{local}$. Local critical temperature of the contact is $T_C^{local} \approx 50$ K. The spectra were shifted vertically for clarity. Andreev minima positions (at T = 4.2 K) for the large gap are marked $n_L$ indexes, for the small gap — as $n_S = 1$.

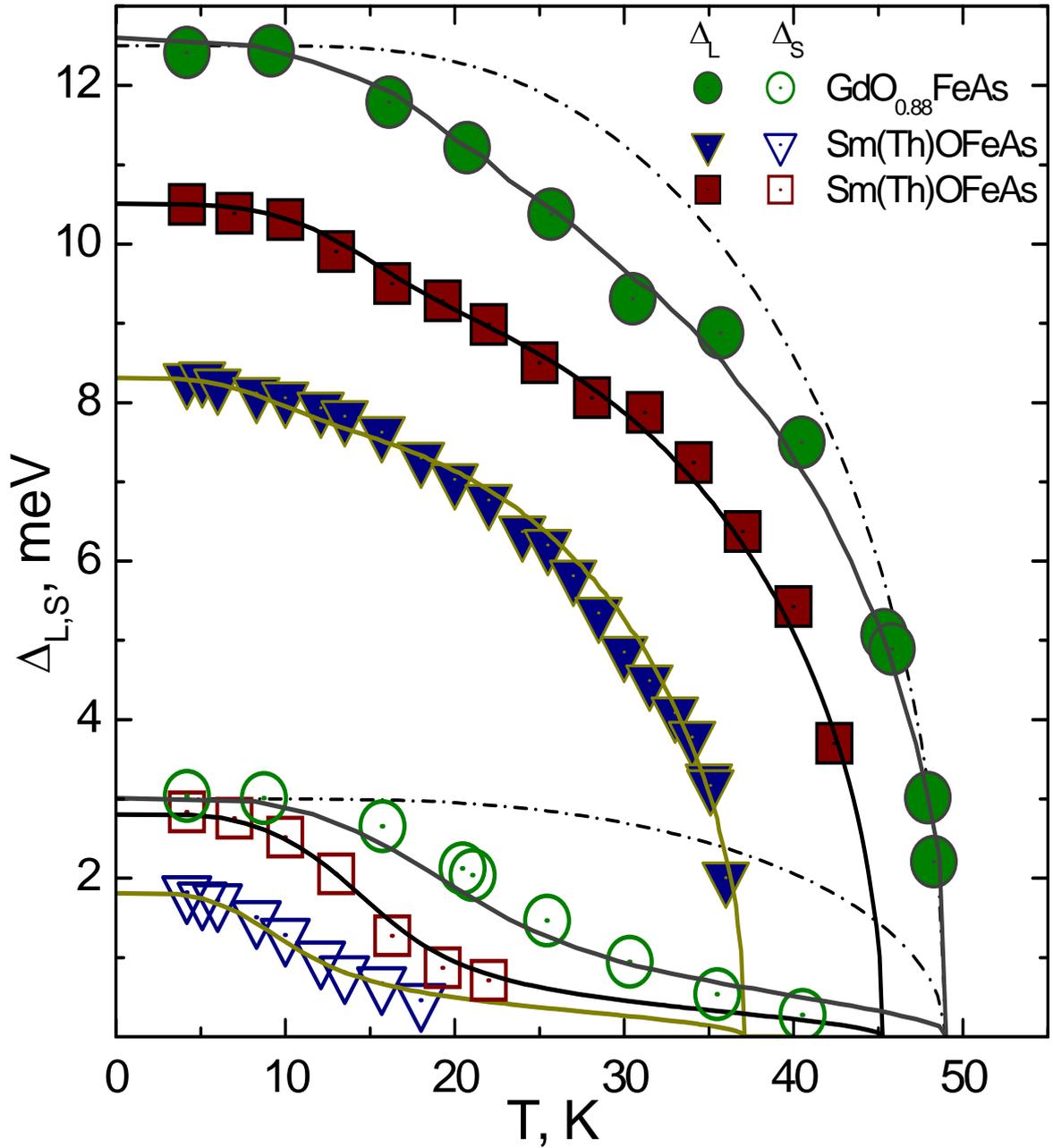

Fig. 4. Temperature dependence for the large gap (solid symbols) and for the small gap (open symbols) for $GdO_{0.88}FeAs$ (circles) with $T_C^{local} \approx 49$ K, Sm(Th)OFeAs with $T_C^{local} \approx 45$ K (squares) and with $T_C^{local} \approx 37$ K (triangles). Theoretical BCS-like functions in the framework of single-gap model (dash-dot lines) and two-gap model (solid lines) are presented for comparison.

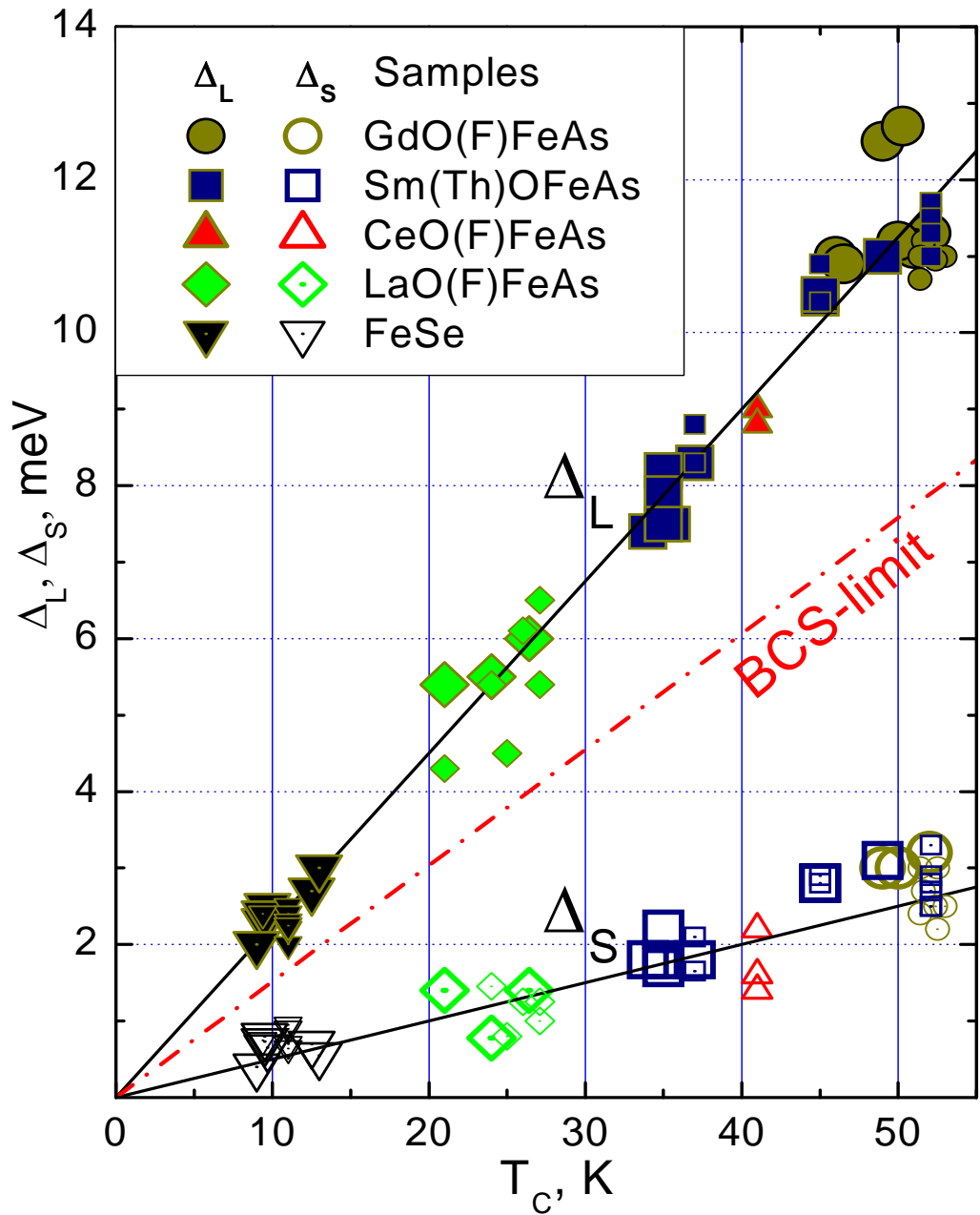

Fig. 5. The dependence of the large gap (solid symbols) and the small gap (open symbols) on the critical temperature for Gd-1111 (circles), Sm-1111 (squares), Ce-1111 (up triangles), La-1111 (rhombs), and FeSe (down triangles). Large symbols depict the data corresponding to $T_C^{local}$, small symbols — those corresponding to $T_C^{bulk}$. Solid guidelines are shown for clarity, the BCS-limit is shown by dash-dot line.

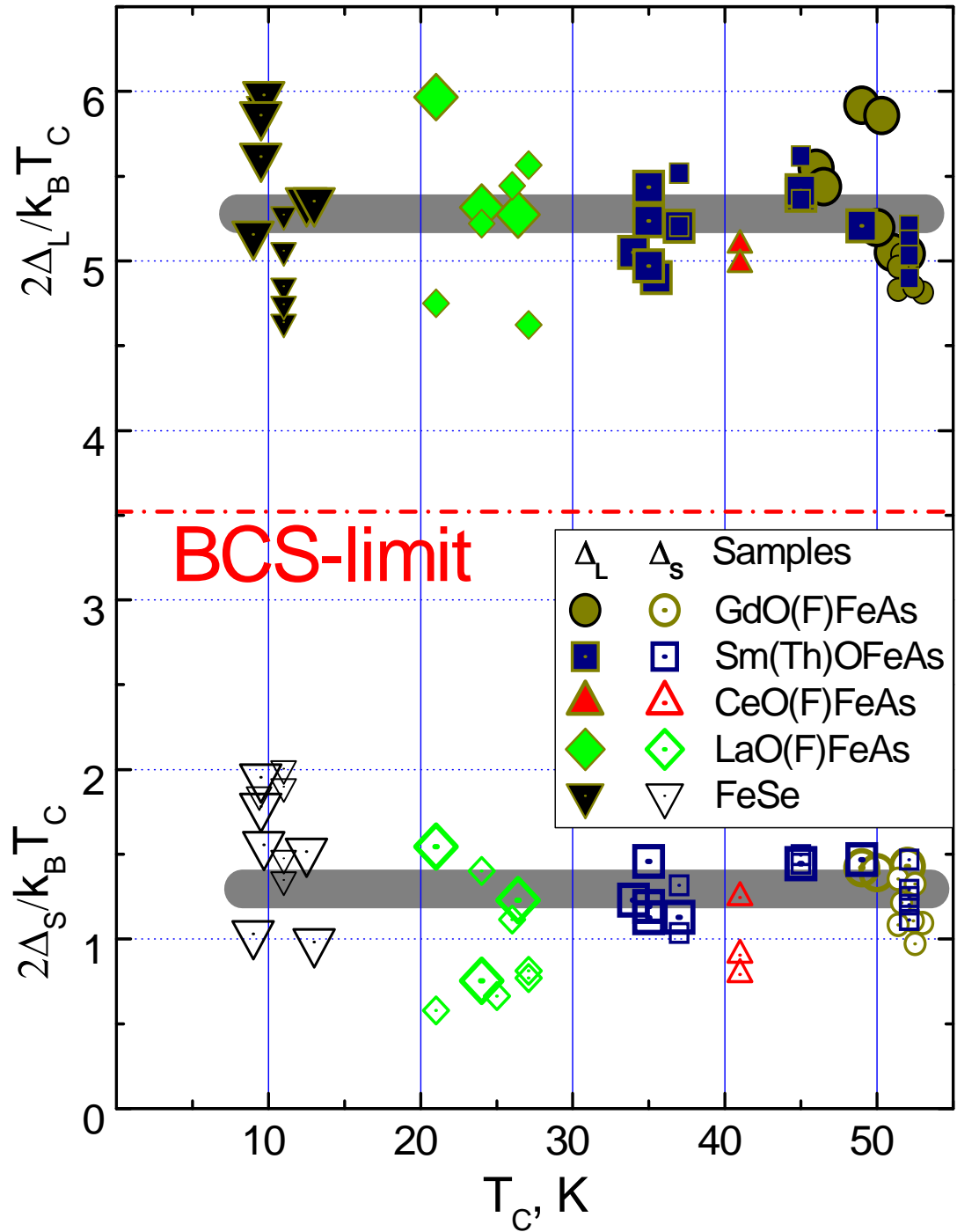

Fig. 6. The dependence of the BCS-ratio $2\Delta_i/k_BT_C$ for the large gap (solid symbols) and the small gap (open symbols) on the critical temperature for Gd-1111 (circles), Sm-1111 (squares), Ce-1111 (up triangles), La-1111 (rhombs), and FeSe (down triangles). Large symbols depict the data corresponding to $T_C^{local}$, small symbols — those corresponding to $T_C^{bulk}$. Solid guidelines show the average BCS-ratios, the BCS-limit is shown by dash-dot line.